# A Hybrid Architecture for Multi-Stage Claim Document Understanding: Combining Vision-Language Models and Machine Learning for Real-Time Processing

Lilu Cheng, Jingjun Lu, Yi Xuan Chan, Quoc Khai Nguyen,

John Bi, Sean Ho

AI Team, Fullerton Health

lilu.cheng@fullertonhealth.com

## Abstract

Claims documents are fundamental to healthcare and insurance operations, serving as the basis for reimbursement, auditing, and compliance. However, these documents are typically not born-digital; they often exist as scanned PDFs or photographs captured under uncontrolled conditions. Consequently, they exhibit significant content heterogeneity—ranging from typed invoices to handwritten medical reports—and linguistic diversity. This challenge is exemplified by operations at Fullerton Health, which handles tens of millions of claims annually across nine markets (including Singapore, the Philippines, Indonesia, Malaysia, Mainland China, Hong Kong, Vietnam, Papua New Guinea, and Cambodia). Such variability, coupled with inconsistent image quality and diverse layouts, poses a significant obstacle to automated parsing and structured information extraction.

This paper presents a robust multi-stage pipeline that integrates the multilingual optical character recognition (OCR) engine PaddleOCR, a traditional Logistic Regression classifier, and a compact Vision–Language Model (VLM), Qwen 2.5-VL-7B, to achieve efficient and accurate field extraction from large-scale claims data. The proposed system achieves a document-type classification accuracy of over 95% and a field-level extraction accuracy of approximately 87%, while maintaining an average processing latency of under 2 seconds per document. Compared to manual processing, which typically requires around 10 minutes per claim, our system delivers a 300× improvement in efficiency. These results demonstrate that combining traditional machine-learning models with modern VLMs enables production-grade accuracy and speed for real-world automation. The solution has been successfully deployed in our mobile application, currently processing tens of thousands of claims weekly from Vietnam and Singapore.

## 1 Introduction

Insurance claim processing remains a labor-intensive and inefficient operation. This stems from the complexity of the workflow, which requires human operators to perform not only document review but also information extraction, policy verification, and final adjudication. Consequently, the processing capacity of individual operators is severely constrained, creating a bottleneck for handling large volumes of claims. Accurate text understanding and information extraction are pivotal yet currently inefficient steps in this loop. This bottleneck impacts not only professional administrators but also claimants. For instance, during mobile claim submissions, users are compelled to manually input specific fields found in their medical records. However, locating and verifying this information on a small mobile interface is challenging, especially given the cognitive gap: users are often unfamiliar with the structure and terminology of heterogeneous documents such as discharge summaries. As a result, the submission process is remarkably inefficient, typically requiring at least ten minutes per claim. In fact, many rule-based approaches have been tried to solve this problem (Goolla, 2025; Kumar and Sharma, 2024). However, the heterogeneity of these documents—different templates, languages, and layouts—deny that effort. Recently, with the large language model has been developed especially large Vision-Language Models (VLMs) that can understand the context in documents. It makes it is possible to understand the document and automate extract the required fields from claim documents. Although Large VLMs possess the advanced semantic understanding required for accurate extraction, they are often impractical for real-world claim processing. The handling of sensitive healthcare data imposes rigid constraints on model selection, ruling out many high-performing but privacy-risky external models. Additionally, the prohibitive inference costs associated with large VLMs hinder scalability. Thus, the challenge lies in bridging the performance gap: achieving the sophisticated understanding of large models within the constraints of compact, privacy-preserving frameworks.

This paper introduces a multi-stage, resource-efficient architecture that leverages compact VLMs and multilingual OCR to extract structured fields from scanned claims (Fig. 1). Our design decomposes the task into pre-processing, hybrid classification, adaptive extraction, and post-processing, each optimized for robustness and interpretability. Figure 1 shows an overview of the proposed architecture.

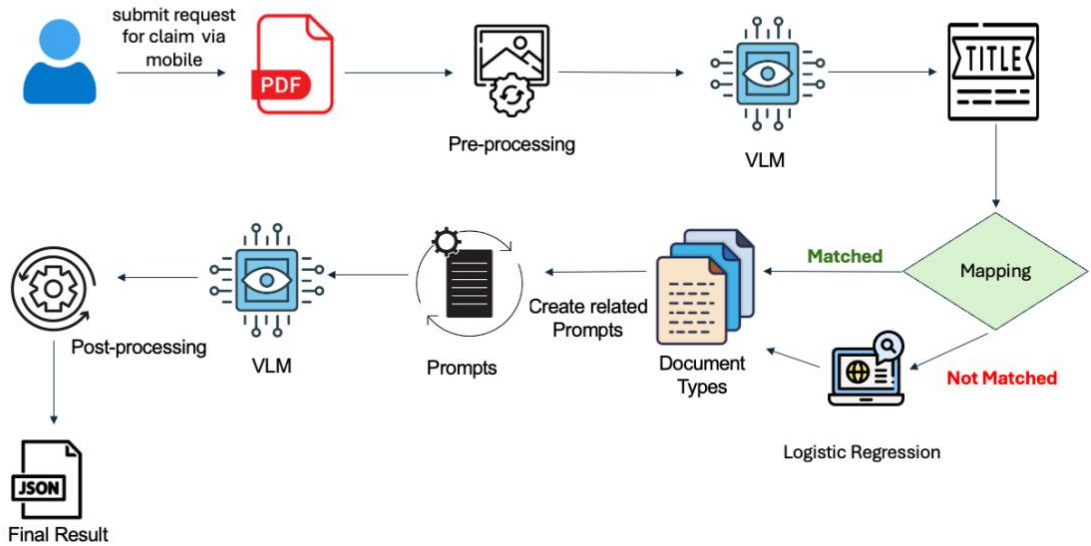

Figure 1 **Architecture of the multi-stage claim document extraction pipeline.** The workflow proceeds as follows: (a) **Pre-processing**: Raw submissions are segmented and rasterized into images. (b) **Hybrid Classification**: A primary VLM extracts document titles for rule-based mapping, with a Logistic Regression model serving as a fallback for unmapped cases. (c) **Adaptive Extraction**: Based on the identified document type, specific prompts are generated to guide the secondary VLM in extracting target fields. (d) **Post-processing**: Extracted entities are normalized via Elasticsearch for knowledge base grounding.

## 2 Background

Research in document understanding generally falls into two core areas: parsing and extraction.

Parsing in OCR primarily involves interpreting document layouts and identifying textual content. Earlier generations of OCR systems (e.g., Casey and Lecolinet, 1996; Mori et al., 1999) relied on manually engineered features and rule-based heuristics, which worked reasonably well under controlled conditions but struggled to cope with the variability and complexity of real-world documents. The emergence of deep learning—especially convolutional neural networks (CNNs) and their subsequent variants—marked a paradigm shift toward data-driven OCR (Goodfellow et al., 2014; Shi et al., 2015). These models brought significant gains in recognition accuracy, robustness, and adaptability. Yet, as AI applications expanded, OCR technologies began facing new demands: they must process a far wider spectrum of inputs, including handwritten text, multilingual documents, rare scripts, and layouts combining tables, charts, and embedded images. In both industry and academia, OCR has evolved from a standalone recognition task into a key enabler for downstream applications such as document understanding, key information extraction (KIE), and semantic retrieval within end-to-end intelligent systems. With the rapid progress of large language models (LLMs) and retrieval-augmented generation (RAG) systems, the role of OCR has become even more crucial. These systems depend on vast amounts of accurate, diverse, and well-structured text data for training and inference. In this context, OCR functions not merely as a text digitization tool but as a foundational component that drives the entire knowledge pipeline—from converting historical archives into searchable text to enabling real-time question answering across multimodal document collections. The precision and completeness of OCR outputs directly affect the reliability and performance of LLM-based applications, particularly in domains dominated by scanned or image-based materials such as legal, academic, or business documents. Modern lightweight OCR frameworks, such as PaddleOCR v3 (Cui et al., 2025), now support multilingual, rotation-aware recognition and hierarchical document parsing, maintaining high accuracy even on low-quality scans. In parallel, layout-aware transformer architectures—including LayoutLM (Xu et al., 2020), LiLT (Yang et al., 2021), and DocLLM (Yao et al., 2024)—have further advanced the field by jointly modeling textual and spatial features to more effectively capture the underlying document structure.

Extraction, in contrast, aims to convert unstructured text into structured data. Traditional approaches relied heavily on rule-based or keyword-matching methods built on parsing results. While effective for clean, single-language documents, these approaches struggle with multilingual content and handwritten inputs where simple keyword matching fails. To overcome these limitations, recent research has leveraged Vision-Language Models (VLMs) capable of understanding the semantic meaning of entire documents. Models such as GPT-4o (OpenAI, 2024) and Qwen-2.5-VL (Bai et al. 2025) have demonstrated remarkable capabilities in semantic comprehension and multimodal reasoning. However, in domains such as healthcare, finance, data privacy constraints often prohibit transmitting sensitive documents to external VLM services, even if large proprietary models like GPT-4o or GPT-o4 offer superior vision capabilities. Consequently, open-source VLMs such as MiniCPM-o 2.6, InternVL2.5-8B, Qwen2.5-VL-7B that can be deployed in private, or on-premises environments present a practical alternative. Among these, Qwen-2.5-VL 7B stands out for its balance between performance and deploy ability—it can be hosted on a single NVIDIA A100 GPU, making it particularly suitable for localized healthcare applications.

## 3 Methodology

Let $\mathcal{T}$ denote the finite set of claim document types (e.g., claim form, invoice, receipt, medical report). Each type $t \in \mathcal{T}$ corresponds to a schema $\mathcal{S}_t = \{f_1(t), f_2(t), \cdots, f_m(t)\}$ that defines the set of fields required for extraction, where each field $f_i(t)$ defines a specific item to extract, such as patient name, policy number.

Let the document instance be denoted as

$$x = \{p_1, p_2, \cdots, p_n\},$$

where each page $p_i$ represents a scanned image or a rendered page from a PDF file. Pages may contain printed, handwritten, or tabular data, possibly in multiple languages. The task is to automatically infer the document type and extract schema-specific fields.

**Document Type Classification**

Identify the document's semantic type:

$$t = g(x), \text{ where } t \in \mathcal{T}.$$

**Schema-conditioned Field Extraction**

For the identified document type $\hat{t}$, extract the corresponding fields defined in $\mathcal{S}_{\hat{t}}$:

$$\hat{y} = h(x, \hat{t}) = \{(f, \hat{v}_f, \hat{b}_f, \hat{c}_f) | f \in \mathcal{S}_{\hat{t}}\}$$

where $\hat{v}_f$ represents the extracted field value, $\hat{b}_f$ represents the evidence (e.g., bounding box or region), and $\hat{c}_f$ the extraction confidence.

The objective is to maximize document-type classification accuracy and field-level extraction quality while minimizing system latency and ensuring interpretability and auditability:

$$\max\left(Acc_{type}(g) + FLA(h)\right) \quad s.t. \min Latency(g, h)$$

To meet these requirements, we propose a four-stage pipeline: (1) Pre-processing, (2) Hybrid Classification, (3) Adaptive Extraction, (4) Post-processing.

## 3.1 Pre-processing

For the claim documents, most inputs are scanned copies or photos. Many of these photos are high-quality, resulting in large file sizes (up to 50 MB), which can significantly slow down the downstream VLM extraction process. Therefore, a pre-processing stage that includes image split and resizing is necessary.

In traditional OCR systems, text recognition often fails when documents are rotated (Goodfellow et al., 2014; Shi et al., 2015). However, since PaddleOCR v3 already has the image quality enhancing and distortion or orientation adjusting system (Cui et al., 2025), it remains robust even when the input image is rotated (as verified in our experiments). Thus, in our pipeline, we will not handle with the orientation issue. Instead, we only two simple pre-processing steps including split and resize:

Split – Load the PDF files and convert each page into an image. These page images are then used as inputs for both the PaddleOCR v3 and Qwen-2.5-VL-7B models.

Resize – The Qwen-2.5-vl 7B model supports a wide range of resolution inputs. By default, it uses the native resolution for input, but higher resolutions can enhance performance at the cost of more computation. Users can set the minimum and maximum number of pixels to achieve an optimal configuration for their needs, such as a token count range of 256-1280, to balance speed and memory usage. In our system, we reduce the large images to a resolution of 1024 pixels:

- Ensures sufficient visual detail for document text, tables and handwritten content without excessive overhead, and
- Aligns with the model's optimal token-budget trade-off, thus delivering good extraction accuracy while controlling latency and memory consumption.

After pre-processing, each page is passed through an OCR system to recover textual content.

We employ PaddleOCR v3 framework, chosen for its strong multilingual coverage, high speed, and robustness in handling noisy or scanned documents. As we mentioned above, PaddleOCR V3 consists of two main components:

Text Detection – A differentiable bounding-box detector identifies text regions across multiple scales. This capability is particularly important for complex financial or claim documents that contain mixed layouts such as narrative text, tables, and marginal notes.

Text Recognition – The detected regions are then transcribed into character sequences using language-specific recognition heads.

It supports more than 80 languages, including English, Simplified Chinese, Vietnam and Traditional Chinese, in both printed and handwritten forms. For each recognized token, the system outputs not only the transcribed text but also a confidence score and the spatial coordinates of its bounding box (Fig. 2). These metadata enable downstream modules to filter out low-confidence results and preserve layout-sensitive structures, such as tables and forms. In our solution we need use confidence score to guide downstream users to pay more attention on the low confidence results.

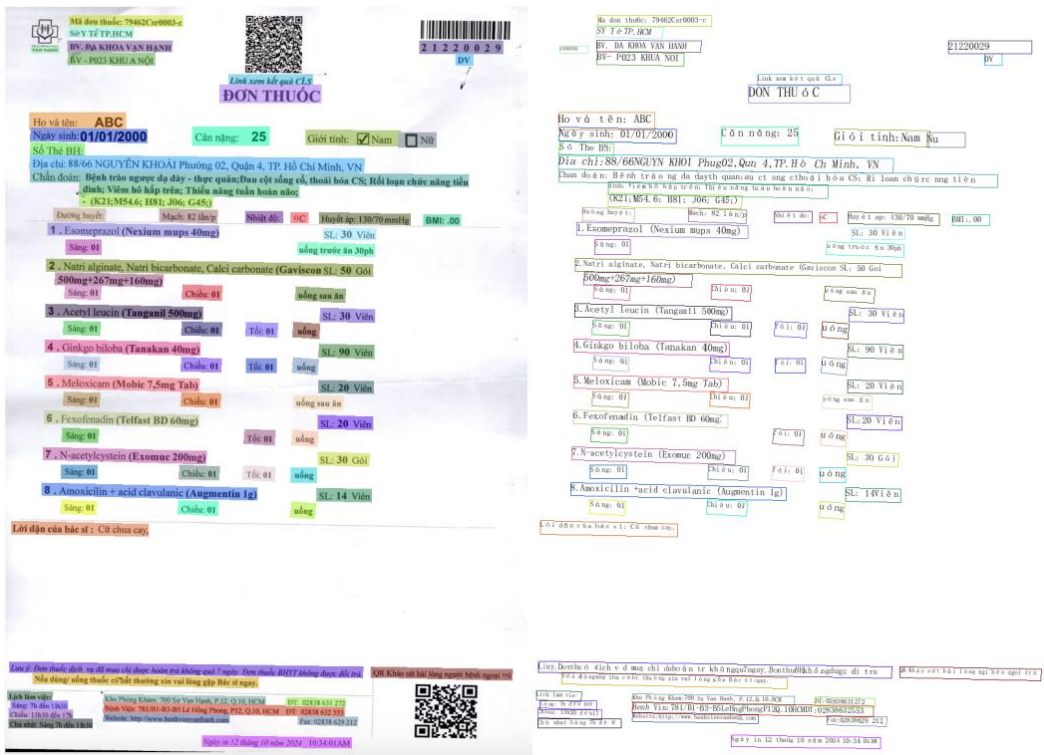

Figure 2 An example of using PaddleOCR on a Vietnam prescription

By leveraging PaddleOCR v3, our pipeline achieves reliable multilingual text transcription and spatial awareness, providing a strong foundation for subsequent extraction and analysis tasks. Most importantly, we can cross-check our results against the outputs from the VLM to enhance overall accuracy.

## 3.2 Hybrid Classification

The claim submission typically comprises more than twenty document types such as claim forms, prescription, receipts and invoice, as summarized in Table 1. Moreover, different insurance companies impose varying documentation requirements. Usually, for the mobile claim application, each uploaded file is treated as a single document type. Although some submissions may include merged files containing multiple document types, those cases are handled by an advanced system that we will introduce in future. Document classification is therefore essential not only for routing each file to the appropriate extraction schema, but also for validating claim completeness. Certain insurers, as well as third-party administrators (TPAs) such as Fullerton Health, require that every claim submission include at least a claim form together with an invoice or receipt. Accurate classification ensures such mandatory documents are present before the claim proceeds to adjudication. In this study, each page within a document is assigned a semantic document type, for example, claim form, invoice, receipt, or medical report, to determine the appropriate schema for subsequent field extraction. To achieve robust document classification, we integrate VLM representations with traditional machine learning features, leveraging both semantic understanding and layout-based cues.

Table 1 Document types of Singapore and Vietnam

|    | Singapore | Vietnam |
|----|-----------|---------|
| 1  | Hospital discharge summary | Hospital discharge summary |
| 2  | Final summary hospital bills | |
| 3  | Final itemized hospital bills | Itemized hospital bills |
| 4  | Medical certificates | Medical certificates |
| 5  | Invoices | Invoices |
| 6  | Receipt | Receipt |
| 7  | Referral letter | |
| 8  | Letter of guarantee | Letter of guarantee |
| 9  | X-ray reports | X-ray reports |
| 10 | Diagnostics test reports | Diagnostics test reports |
| 11 | Lab reports | Lab reports |
| 12 | Prescriptions | Prescriptions |
| 13 | Histology reports | Histology reports |
| 14 | CPF statements | |
| 15 | Claim settlement | Claim settlement |
| 16 | Guarantee letter (pre-admission) request forms | Guarantee letter (pre-admission) request forms |
| 17 | Test order form | Test order form |
| 18 | Hospital pre-admission form | Hospital pre-admission form |
| 19 | Claim form | Claim form |
| 20 | Initial guarantee letters | Initial guarantee letters |
| 21 | Final guarantee letters | Final guarantee letters |
| 22 | | Discharge certificates |
| 23 | | Surgery certificates |
| 24 | | Birth Certificates |
| 25 | | Record of physiotherapy |
| 26 | | Accident reports |
| 27 | | Vehicle registration |
| 28 | | Driver license |
| 29 | | National id |
| 30 | | Dental treatment form |

VLM-based classification: Each page image is first processed by the Qwen-2.5-VL-7B model using the prompt "Identify the document type: claim form, invoice, medical report, or receipt." With a definition of those document type. The model leverages layout and visual cues (e.g., titles, logos, and structured forms) to infer form titles. In our experiments, directly using the VLM for full document-type classification yielded only around 70% accuracy, which is lower than that of conventional machine learning models which is 76%. The main limitation stems from the compact capacity of the deployed VLM version—document type classification requires broader contextual reasoning than the model can provide. Consequently, we employ the VLM primarily to extract form titles, which often contain strong semantic hints such as Referral Letter or Discharge Summary. The model performs significantly better on this extraction task because title regions are usually short, well-defined text segments that align closely with the VLM's natural language understanding and visual grounding capabilities. By focusing on localized text-image relationships rather than full-page reasoning, the model can more accurately interpret font size, layout position, and linguistic patterns associated with document headers. In most cases, this approach achieves more than 95% accuracy; however, titles from invoices or bills can still be ambiguous and visually similar, making them more challenging to distinguish.

Machine learning–based classification: to address these cases, we train a traditional machine learning classifier using logistic regression with TF-IDF embedding as input. The 2000 samples are classified into training set and testing set with ratio of 80:20. We compared several models, including Random Forest, XGBoost, and Logistic Regression, under identical feature inputs. Empirically, logistic regression achieved the best balance of accuracy and generalization and is therefore adopted as our final document-type classifier (Table 2).

Table 2 Traditional machine learning model performance

| Machine learning model | Accuracy (%) |
|---|---|
| XGBoost | 76 |
| Random Forest | 77 |
| Logistic Regression | 87 |

## 3.3 Adaptive Extraction via Compact VLM

After classification, the system triggers schema-specific field extraction using Qwen-2.5-VL-7B. Each schema defines a fixed set of target fields. For example:

*Claim Form: Claim id, Patient Name, Policy Number, Claim Amount*

*Invoice: Provider, Date, Total Amount,*

*Medical Report: Diagnosis, Provider, Doctor's Name, Admission Date*

*Receipt: Receipt Number, Provider, Paid Amount, Payment Date*

A prompt generator dynamically constructs natural-language instructions for the model, such as: "Extract the following fields from this invoice image: Provider Name, Date of Service, and Total Amount. Return the result as JSON."

In practice, we find that using a one-slot structured prompt—composed of the following four components—substantially improves extraction accuracy:

Role Definition – Specifies the model's function or perspective (e.g., "You are an information extraction assistant…").

Field Definition – Clearly describes the target fields and their semantic boundaries.

Output Format Specification – Defines the expected data schema and output style (e.g., JSON or key–value pairs).

Example of Expected Output – Provides a reference example to guide the model's response structure and tone.

The compact VLM jointly processes both visual layouts and OCR tokens, enabling it to infer semantic relationships between field labels and corresponding values. The model outputs field values along with bounding boxes and confidence scores which are provided by PaddleOCR for spatial traceability.

Compared with large-scale VLMs, the compact version can be hosted on single GPU server such as H100 while maintaining competitive accuracy.

The extracted information is exported in JSON format, as illustrated below:

```json
{
 "claim_id": "C2024-0001",
 "patient_name": "ABC",
 "policy_no": "VN111",
 "diagnosis": "Acute bronchitis",
 "provider": "Hanoi General Hospital",
  "visit_date": "2024-10-05",
  "total_amount": 1650000,
 }
```

### 3.4 Post-processing

Although the VLM produces highly accurate results, certain errors may still occur. For instance, hospital names can sometimes be inconsistent or incorrectly extracted. To address this, we apply an Elasticsearch-based normalization module (Gormley and Tong, 2015), which replaces each extracted hospital name with the most similar entry from a predefined reference list. To facilitate this process, we maintain a pre-stored database of hospital names collected from verified sources such as the Ministry of Health, insurance provider networks, and our own internal system records. During post-processing, each extracted name is compared against this repository using fuzzy string matching and similarity scoring. The system then substitutes low-confidence or ambiguous results with the closest standardized hospital name. This ensures consistency across documents and significantly improves the accuracy of downstream tasks such as entity linking, claims matching, and analytics reporting.

Additionally, date formats are sometimes unstable. While the expected format is DD/MM/YYYY, occasional outputs include extra time components (e.g., DD/MM/YYYY HH:MM). A date-format validation rule is therefore applied to enforce consistent formatting across all outputs.

## 4 Results

We evaluated the proposed pipeline on 4,200 multilingual claim documents (approximately 32,000 pages) collected from Singapore and Vietnam, covering more than 20 document types such as claim forms, invoices, medical reports, and receipts (Table 1). The dataset consists of English (50%) and Vietnamese (50%) documents, with both printed (70%) and handwritten (30%) content. Evaluation metrics comprise document-type classification accuracy, field-level extraction accuracy, average latency per document. As summarized in Table 3, the proposed method achieves 87 % field-level accuracy and 97 % classification accuracy. Moreover, average processing latency per document is to 2 s, demonstrating a substantial improvement in computational efficiency. In comparison, manual processing typically takes around 10 minutes per document, meaning our system delivers more than a 300× improvement in efficiency.

Table 3 Results Experimental Results on Singapore and Vietnam Claim Datasets

|  | Classification Accuracy(%) | Extraction Accuracy(%) | Average process time per document(s) |
|---|---|---|---|
| SG | 93 | 87 | 1.8 |
| VN | 97 | 75 | 2 |

Despite these strong results, several error sources remain.

First, OCR noise from low-contrast or handwritten text can lead to token loss or misrecognition. For example, in some Vietnamese claim forms that use dotted lines, handwritten entries—especially for the claim amount—are often misread. The simplest solution is to provide blank forms for customers to fill in, thereby improving text clarity and OCR accuracy.

Second, date ambiguity remains a challenge. When multiple dates (e.g., visit date and treatment date) appear in the same document, the VLM occasionally selects the wrong one. In addition, compact models still show limited stability in distinguishing between very similar terms.

With the achieved accuracy, we integrated our AI solution into the claim submission application. As users upload their claim documents, the system automatically determines the document types and extracts the key fields required by the claim policies, populating the application form in real time. The interface remains user-friendly, enabling users to review and correct any residual errors identified in the earlier analysis. As a result, the system saves users hundreds of thousands of minutes per week, significantly improving the overall efficiency of the claims submission process.

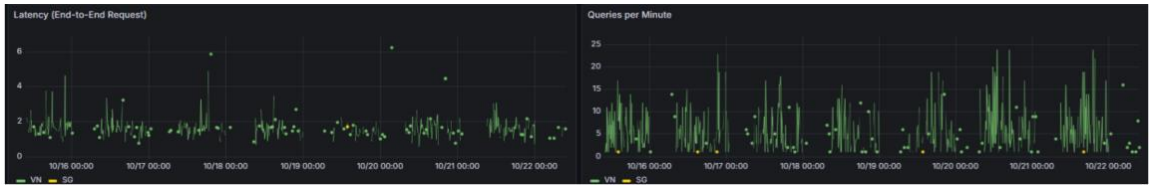

Figure 3 Screenshot of OCR API Call Summary. The left panel shows the response time, with most requests completing in under two seconds. Few longer response times are likely due to larger document sizes. The right panel illustrates the number of API calls over time, providing an overview of the system's usage and performance trends.

## 5 Conclusion and Future Works

We presented a multi-stage Vision–Language Model pipeline for comprehensive claims document understanding for real-time processing. The proposed system demonstrates high accuracy, low latency, strong scalability, and interpretability, utilizing compact models that can be efficiently deployed on a single GPU server such as an H100 or A100. This solution has been fully integrated into our production claims system as an API, serving tens of thousands of users daily and delivering accurate results within seconds (Fig. 3). In practice, our AI solution saves hundreds of thousands of minutes per week, reducing manual processing time from 10 minutes to approximately two seconds per document.

Looking ahead, we plan to extend the system's multilingual capabilities to include additional languages such as Simplified Chinese, Traditional Chinese, Malay, and Indonesian, as well as support for claim documents from more countries. Ultimately, our goal is to scale this solution to serve our millions of users. As document understanding constitutes a foundational step toward fully automated claim processing, we plan to build a multi-agent claim adjudication framework that leverages the outputs of the proposed multi-stage field extraction pipeline.

**Acknowledgments**

This work was developed by the Fullerton Health AI Team in collaboration with regional claims operations, product teams and engineer team across Singapore and Vietnam. We extend our sincere appreciation to Dr. Morrison Loh, Mr. Ho Kuen Loon, and Mr. Brian Lim for their invaluable guidance, encouragement, and support throughout this work.